# Overview of Spanish Tracking R&D for FLC

Alberto Ruiz-Jimeno
(on behalf of the Spanish Network for Future Linear Accelerators)

IFCA, Instituto de Física de Cantabria ( CSIC-Universidad de Cantabria)
Avda. los Castros, s/n
39005 Santander- Spain

An overview of the present and foreseen R&D activities of the Spanish network for future accelerators aiming to participate in the design and construction of the forward tracker and vertex detectors of the Future Linear Colliders, is shown.

## 1 Introduction

Several Spanish high energy institutions have joined in a coordinated effort of R&D for future linear accelerators and detectors for particle physics. As it is shown in the Figure 1, the network covers a big part of the national geography. Most of the groups are members of big collaborations, as SiLC (Silicon for Large Colliders [1]), DEPFET (Depleted p-Channel Field Effect Transistor [2]) or CALICE (Calorimetry for an ILC detector [3]), and participate in old European Projects EUDET (Detector R&D towards the International Linear Collider[4])... or new ones as AIDA (Advanced European Infrastructures for Detectors and Accelerators [5]).

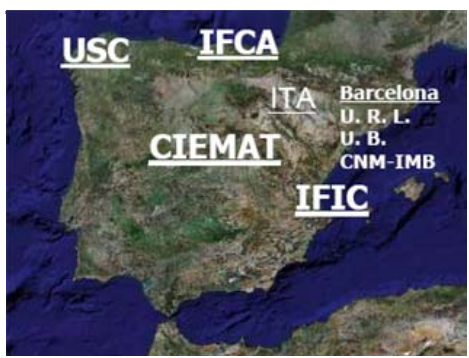

Figure 1: Spanish Institutions for R&D For Future Linear Accelerators and Detectors

The Spanish interest and evolution of its activities are driven by the present participation in the LHC experiments, its upgrade projects for SLHC and its participation, as DEPFET members, in the vertex detector for Belle-II.

Concerning the detector development activities, the long-term goal is to participate in the design and construction of the forward tracker and vertex detectors of the Future Linear Colliders. Its main related activities are the research and development of technologies to reach accurate and efficient reconstruction of charged particle trajectories as well as primary and secondary vertexes, alignment, integration studies, simulation and optimization.

## 2 The end cap tracking physics case

We are doing an evaluation of the physics case for the end cap tracking and the most challenging aspects which will determine the detector design for this zone.

The paper [6] gives a little guidance for forward detector design from standard benchmark reactions ($\cos \theta < 0.95$). At a high energy e+e- collider several potentially very interesting physics analysis require excellent tracking and vertexing performance. These arguments become more urgent as the center of mass energy increases. Precise electron reconstruction is of particular importance due to the energy loss; single electron samples generated by us show a loss of 3 GeV/c in average for the end cap electrons and 150-200 MeV/c for central ones (Figure 2)

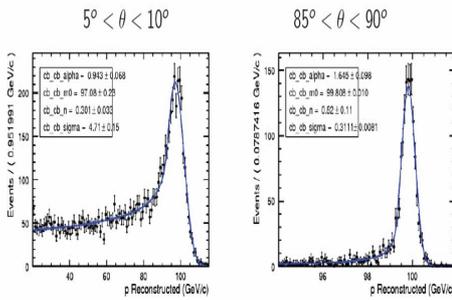

Figure 2: Electron energy loss

There is significant physics to be gained or lost in the end cap region. If the central vertexing performance is somewhat of a challenge, maintaining good performance at small polar angle is much more difficult.

A simple-minded layout optimization of the Vertex and Forward-tracker indicates that the better performance is reached by minimizing the $z_{gap}$ between the vertex and the first disk of the forward tracker and choosing an appropriate geometry considering the routing for the services and cables. We are doing detailed studies of the optimal geometry.

As an example, using a toy geometry, in Figure 3 it is shown how the tracking performance is degraded when the services routing goes along the beam pipe versus going upward.

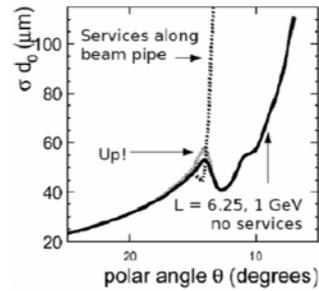

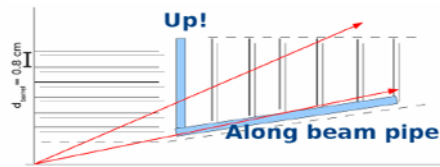

Figure 3: Impact parameter resolution for a toy geometry of the tracker

## 3 Studies of technologies for the vertex

Different technologies for the vertex detector pixels are being studied by the Spanish groups, including DEPFET, Single-APDs and Active pixels. It exists also contacts with experts on other technologies which are ILC candidates, as MAPS, FPCCD,...

3.1 DEPFET

DEPFET technology [7] is one of the most advanced candidates for ILC vertex detector experiments. It has been developed by the MPI Semiconductor Laboratory, in Munich.

Spanish groups leaded by the IFIC (Instituto de Física Corpuscular, CSIC-Univ. Valencia) have been involved since long ago in the DEPFET collaboration, participating in the design, tests and mechanics. Belle-2 has also selected this technology as baseline for the vertex detector. Some of the present activities are shown in Figure 4.

In DEPFET, each pixel is a p-channel Field Effect Transistor integrated on a completely depleted bulk. An additional deep n-implant creates a potential minimum for electrons, considered as an internal gate, where they are accumulated, modulating the transistor current. The charge accumulated can be removed by a clear contact placed in the periphery of each pixel.

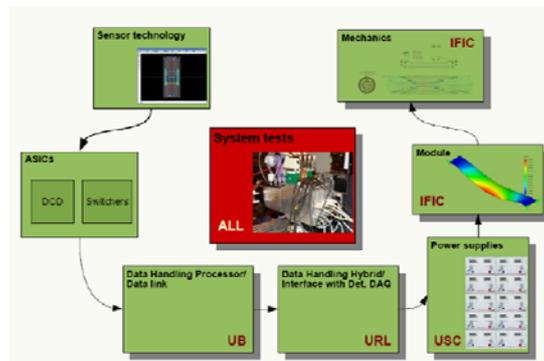

Figure 4: Spanish activities in DEPFET

DEPFET has a good resolution and sensitivity, low power consumption, low material budget and good signal over noise ratio. It is a monolithic sensor technology with the electronics built on the pixel itself. It can be thinned out to 50 microns. For ILC the goal is to reach a readout per row of about 50 nsec and a noise of about 40 electrons at high bandwidth, which can be reached by a small capacitance and first in-pixel amplification.

An intense beam test programme has been developed over the last years to demonstrate that the concept will be able to cover the requirements of high precision and transparency required for the ILC vertex detectors. Spanish groups have been very much involved on it. Intrinsic resolutions at the level of 1.2 microns have been already achieved.

Spanish groups are interested in the evaluation of DEPFET sensors, both as single pixel structures, in order to test the different parameters driving the intrinsic performance of the sensors, as well as pixel array structures, including final read-out electronics and control.

### 3.2 APDs

We are interested in developing CMOS APDs [8] technology as a promising one for ILC. These monolithic devices give a very fast signal, but have detector instabilities (dark current, afterpulsing noise and cross-talks) which contribute to the detector response, having a deep impact on the readout details. We are trying to understand their origin and reduce their incidence, both from simulation and design of pixels and readout structures. It is a challenge particularly for large matrices, as needed for future colliders. On the other side they require basic readout electronics which impact on lower power consumption.

UB, URL and CNM-IMB have a big expertise on this type of devices. We have learn from two different technologies (STM low voltage 130 nm. and AMS High voltage 350 nm.) with optical generation of pairs. The requirements of future linear colliders in terms of charge, speed and area coverage is a real challenge which we will study both in planar and "3D" technologies.

### 3.3 Active Pixels

As one alternative to monolithic pixel sensors, USC is interested in studying active pixel detectors which have a faster signal and read out, allowing time stamping. We are studying the feasibility to instrument pixel sensors with Timepix [9] readout chip and the possibility to thin both the sensor substrate and the readout chip, in order to reduce the amount of material.

## 4 Sensor developments for the end cap tracker

The end cap tracker will consist of about 3 internal disks (< 0.25% X0 radiation lengths) and four external disks which more relaxed conditions of material budget (< 0.65% X0 radiation lengths). The aim is to have a spatial resolution of 7 microns in R$\phi$.

Both pixel (for the inner disks) and micro-strip (with a pitch of ~ 40 μm for the external ones) silicon technology are considered. In both cases a low power Front End Electronics is required. Thinned silicon microstrip detectors for the external disks are needed.

A highly compact mixed-signal chip for readout of silicon strip sensors is being developed by the UB in collaboration with the LPNHE in the frame of the SiLC collaboration. The ASIC will be fabricated in a deep-submicron technology in order to reduce the wasted space. It will contain an analog part, responsible of the amplification, shaping and digital conversion. The chip functioning is fully digital controlled. All the parameters such as bias voltage and current, sparsifier thresholds, event tag and time tag generation, internal calibration system, shaping time frequency and sampling frequencies will be programmable. This ensures a high degree of fault tolerance and high flexibility and robustness.

The characterization of the sensors is made in the laboratory. We are doing studies of procedures to produce edgeless sensors to reduce dead zones as well as removal of pitch adapter to increase the signal over noise.

## 5 Mechanics, alignment and integration studies

The Spanish groups activities cover studies of readout electronics, as well as connectivity, integration, mechanical and thermal monitoring, and alignment. An sketch of some of the present activities is shown in Figure 5.

5.1 Alignment

Silicon sensors currently used as tracking detectors in high energy physics experiments have a weak absorption of infrared light, but enough to use laser beams as pseudo-tracks traversing consecutive sensors. We propose a hybrid alignment system using a part of the silicon tracking detectors, with minor modifications aiming to make them highly transparent to infrared light [10]. Recently, we have done a validation of the optical simulation software with planar multilayered samples and layered diffraction grating and we are doing a measurement of optical test structures and actual sensors produced in the CNM-IMB (see Figure 6).

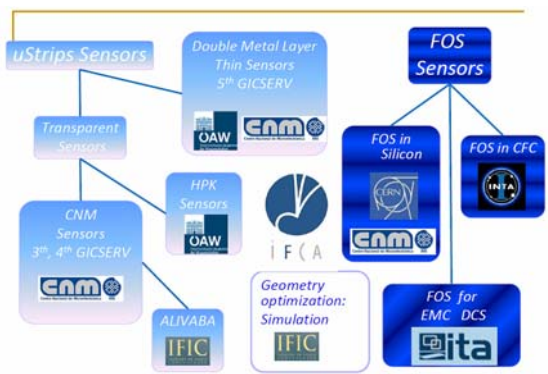

Figure 5: Some Spanish activities on silicon sensors

We have taken a great advantage from the GICSERV program of CNM-IMB which allows to access the infrastructure of the Institute. We achieved transparencies in the order of 70-80% for the test structures and we hope to reach results of the same order for the sensors. Direct comparison with interferometric measurement at the laboratory give accuracies better than 1 μm.

Test beam at CERN SPS has been done in order to compare between track-based and laser alignment as part of the EUDET-SiTRA alignment task. The analysis is still in progress.

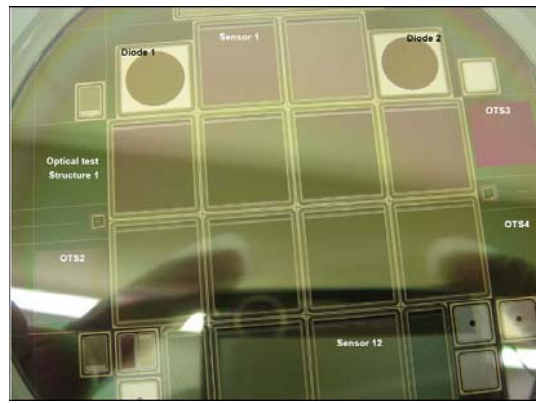

Figure 6: A wafer produced by CNM-IMB for studies of the optical alignment sensors

5.2 R&D on mechanics

We propose to use Fiber Optical Sensors (FOS) as Bragg grating optical transducer (Figure 7) to measure strain and temperature. FOS have good qualities to be used in Future Linear Colliders environment as light-weight, flexible, low thermal conductivity, non-interfering, low-loss and long-range signal transmission. They have also multiplexing capability, can be embedded in composite materials; wavelength is encoded making measurements transferable and neutrals to intensity drifts. They are durable to high strain and support high and low temperatures.

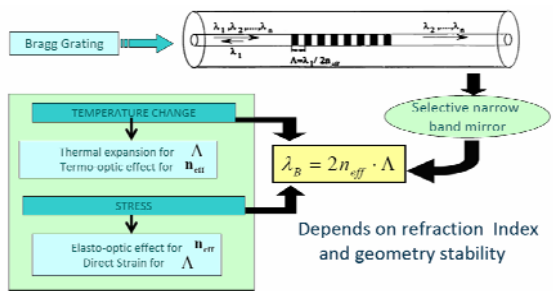

Bare optical fibers are quite inhomogeneous and external agents or radiation can induce changes of the mechanical properties, changing the sensor grating period $\Lambda$. Calibration with different coatings, as acrylate, polyamide... as well as without coatings is one of the activities being done at IFCA with a test setup which allows for redundant calibration.

Figure 7: FOS scheme

A joint project of IFCA in collaboration with the Spanish Aerospace Agency (INTA) will allow irradiation of embedded fibers with different coatings. Also small samples 15x3x3 mm3 of composite laminates with different stack configurations with and without optical fiber embedded is in preparation. Nano indentation characterization of the different components of the samples (coating, cladding) and composite matrix will allow extraction of the mechanical parameters, as poisson, young modulus...

A collaboration with IFIC has been established in order to design for a temperature and position monitor for the PXD of Belle-2.

We are also working towards an engineering design for the ILD Forward Tracker (Figure 8)

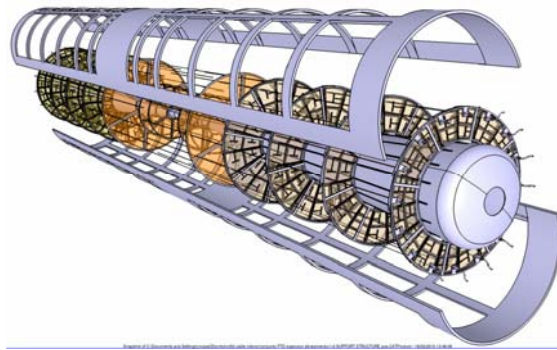

Figure 8: Naif preliminary design of the FTD of ILD

## 6

## 7 Acknowledgments

The reported work was funded under MICINN projects CSD2007-0042 and FPA2008-03564-E/FPA.

## 8 Bibliography

## 9 References

[1] Silicon Tracking for Linear Collider webpage: http://lpnhe-lc.in2p3.fr
[2] Depleted p-Channel Field Effect Transistor Collaboration webpage: http://www.depfet.org
[3]Calorimetry for an ILC detector webpage: https://twiki.cern.ch/twiki/bin/view/CALICE/WebHome
[4] Detector R&D towards the International Linear Collider web page: http://www.eudet.org
[5] Advanced European Infraestructures for Detectors and Accelerators. Project submitted to the FP7-INFRASTRUCTURES-2010-1 call from the European Commission: https://espace.cern.ch/aida/default.aspx
[6] J. Fuster, S. Heinemeyer, C. Lacasta, C. Marinas, A. Ruiz-Jimeno, M. Vos, Forward tracking at the next e+ e- collider part I: the physics case. JINST 4 P08002 (2009).
[7] J. Kemmer and G. Lutz. New Semiconductor Detector Concepts. Nucl. Instr. Met. Phys. Res. A, vol A253, 356-377 (1987).
[8] C.J.Stapels, Nucl. Inst. Met. Phys. Res. A, vol. 579, 94 (2007)
[9] X. Llopart et al., Nucl. Inst. Met. Phys. Res. A, vol. 581, 485-494 (2007)
[10] A. Ruiz et al., Proceedings of the LCWS08 and ILC08, International Linear Collider Workshop 2008,Univ. of Illinois at Chicago (USA), Nov.16-20, 2008